# Excited Rydberg States in TMD Heterostructures

*Jacob J.S. Viner [1], Liam P. McDonnell [1], David A. Ruiz-Tijerina[2], Pasqual Rivera [3], Xiaodong. Xu[3], Vladimir I. Fal'ko [4,5], David C. Smith [1*]*

1 School of Physics and Astronomy, University of Southampton, Southampton SO17 1BJ, United Kingdom.

2 Instituto de Física, Universidad Nacional Autónoma de México. Apartado Postal 20-364, Ciudad de México, 01000, México

3 Department of Physics, University of Washington, Seattle, WA, USA

[4]National Graphene Institute, University of Manchester, M13 9PL, United Kingdom.

[5] Henry Royce Institute for Advanced Materials, University of Manchester, Manchester, M13 9PL, United Kingdom.

**The functional form of Coulomb interactions in the transition metal dichalcogenides and other van der Waals solids is critical to many of their unique properties, e.g. strongly-correlated electron states, superconductivity and emergent ferromagnetism. This paper presents measurements of key excitonic energy levels in $MoSe_2/WSe_2$ heterostructures. These measurements are obtained from resonance Raman experiments on specific Raman peaks only observed at excited states of the excitons. This data is used to validate a model of the Coulomb potential in these structures which predicts the exciton energies to within ~5 meV / 2.5%. This model is used to determine the effect of heterostructure formation on the single-particle band gaps of the layers and will have a wide applicability in designing the next generation of more complex transition metal dichalcogenide structures.**

One of the key reasons for the unique properties of van der Waal's structures is that the reduction of dielectric screening in these 2D systems leads to enhanced Coulomb interactions with functional form quite different from the standard 3D Coulomb potential[1,2]. This Coulomb potential leads to the renormalisation of the single particle electronic energy and an increase in the band gap[3,4]; exciton states which follow a non-hydrogenic series[5] with large oscillator strengths and short radiative lifetimes[6–8]; and stable multi-particle states such as trions[9–12]. It is critical to the existence of strongly-correlated states leading to optically accessible incompressible electron states, superconductivity[13,14], and emergent ferromagnetism[15]. It is responsible for the existence of interlayer excitons in transition metal dichalcogenide (TMD) heterostructures and their mutual repulsion leading to a density dependent blue shift of their emission[16]. The exchange component of the Coulomb interaction is responsible for mixing of the A and B excitons with implications for spintronics[17].

The importance of quantifying the Coulomb interaction in TMD monolayers has led to a series of experiments aimed at measuring the energy of the bright, zero orbital angular momentum (s), excitonic states in these structures using a variety of techniques including reflectivity[5,18,19], photoluminescence excitation[6,20], and magneto-PL[21–23]. In addition two-photon absorption measurements have been performed to allow the energy of the one angular momentum (p) states to be determined[24]. However, despite the wider range of exciting physics, e.g. interlayer excitons, which can be accessed in TMD heterostructures, there are no reliable measurements of the energies of intralayer (single layer) excitons beyond the minimum energy 1s state in such structures. In this paper we discuss the determination of the energy of the (intralayer) A2s and B2s states in a $MoSe_2/WSe_2$ bilayer heterostructure with accuracies of a few meV in most cases. The method relies on unique Raman peaks only observed at 2s transitions which enable the unequivocal assignment of specific spectral features to these transitions. Additionally, we demonstrate that it is possible to theoretically reproduce these energies to within the errors of the experiments, i.e. 2.5%, using a model whose parameters can be calibrated with monolayer exciton energies and constrained by separate experiments, and properly takes into account the distance dependent screening.

The results presented here are part of a larger study of high quality hBN encapsulated MoSe$_2$ and WSe$_2$ monolayers[8] and MoSe$_2$/WSe$_2$ heterobilayers[25,26], with twist angles 2, 6, 20, 57 and 60°, using resonance Raman and reflectivity spectroscopy. All measurements were performed with the samples held at 4 K in the low excitation intensity regime (< 100 µW). The resonance Raman studies were performed with excitation photon energies from 1.6 to 2.27 eV allowing us to probe the MoSe$_2$ A1s, A2s, B1s and B2s and WSe$_2$ A1s, A2s and B1s excitonic resonances. The Raman scattering intensity has been calibrated using absolute scattering rates for the silicon peak at 520 cm$^{-1}$ and corrected for Fabry-Perot effects[27,28].

Colour plots of the Raman results for MoSe$_2$ and WSe$_2$ monolayer and two MoSe$_2$/WSe$_2$ twisted heterostructures, with twist angles of 57 and 20 °, for excitation energies spanning the MoSe$_2$ A1s and B1s intralayer excitons and the WSe$_2$ A1s intralayer excitons are presented in Figs 1, 2 and 3 respectively along with representative Raman spectra at key excitation energies.

As previously observed, the Raman spectra of the samples present a large number of distinct Raman peaks[8,29,30]; 18 in MoSe$_2$ and 16 in WSe$_2$. The Raman peaks show clear resonance behaviour with significant enhancements associated with either the incoming or outgoing photons being at the same energy as one of the bright excitonic states in the two materials. The energy of the two conditions for each of the 2s excitons are presented on the figures using a solid line for the incoming resonance and a dashed line for the outgoing resonance. Despite the Raman scattering at the A1s and B1s excitonic resonances being approximately two orders of magnitude stronger than at the A2s and B2s, key phonons which are only observed at these 2s resonances[8] allow them to be clearly identified. In the case of MoSe$_2$ the 2s specific peaks are at 480, 530 and 580 cm$^{-1}$ (see Figs 1 and 2 and reference[8]). As shown in the colour plots in Figures 1 and 2, the peaks have a clear outgoing resonance with the A2s state and incoming and outgoing resonances with the B2s states. In the case of WSe$_2$ there is a unique Raman feature which is only observed at the A2s[8]. This feature is a peak with a characteristic dispersion/resonance behaviour. It starts as a single peak at 495 cm$^{-1}$ at the incoming resonance energy with intensity that drops to zero for energies just above. It reappears when the laser is tuned to approximately half way between the incoming and outgoing resonances. The peak then disperses to lower shift by 2.5 cm$^{-1}$ with increasing laser energy until excitation energies close to the outgoing resonance at which point its magnitude increases significantly and it splits into three closely spaced peaks; see SI for more details. There are no differences in the behaviour of the 2s specific peaks between heterostructures and monolayers that cannot be explained by a change in the energy or width of the excitonic transitions.

Now that we have identified the key Raman peaks, we can extract their resonance profiles and fit them to obtain the energy and width of the transitions associated with the excitons. The resonance profiles of the peaks associated with the MoSe$_2$ A2s resonance (480, 530 and 580 cm$^{-1}$) in the heterostructures were fitted to a Lorentzian lineshape at the outgoing resonance energy. The energy of the phonon was then subtracted to give the energy of the excitonic state. In the case of the MoSe$_2$ B1s resonance the standard third order perturbation prediction (Shown in Supplementary Section S3) for the Raman scattering was used to fit both incoming and outgoing resonances. For the WSe$_2$ A2s, the outgoing resonance was fitted using a Lorentzian lineshape; the complexity of this resonance (Shown in Supplementary Section S5) makes fitting both incoming and outgoing resonances with a third order perturbation model impossible and the outgoing resonance fit is simpler. The energies for the various exciton states obtained from these fits and fits to the resonance profiles of the $A_1'$ phonon at the A1s and B1s excitons are summarised in Tables 1 and 2; the linewidths and a brief discussion of them is given in the SI. In Table 3 we present results for the difference in energy between different excitonic states in monolayers of MoSe$_2$ and WSe$_2$ from the literature. Where comparison is possible, the results from the current resonance Raman experiments and the literature monolayer results are in agreement to within experimental error.

As presented in Tables 1 and 2, the energies of the intralayer exciton states and the 1s-2s separations are both significantly reduced in the heterostructures. As the binding energy of the excitons scale with the 1s-2s separation[3] we can conclude that the binding energy of the excitons is reduced in the heterostructures. This fact combined with the reduction in the absolute energy of the A1s excitons means we can also conclude that the single particle band gaps of the heterostructures must be reduced and by more than the binding energy of the excitons. In addition to the differences between monolayers and heterostructures there are clearly smaller differences in the energies of the various states in the two heterostructures. Whilst it would be tempting to analyse these in terms of twist dependent hybridisation effects, we feel that the possibility of sample to sample variations, independent of twist angle, means it would be premature to do this at this point.

In order to go further in our interpretation of the results we require a theory for the excitonic energies in the heterostructure. Our model is based upon that set out in[31]. We model the MoSe$_2$ and WSe$_2$ layers as infinitesimally thin films possessing only in-plane electric polarizabilities $\kappa_{MoSe_2}$ and $\kappa_{WSe_2}$, respectively, separated by an interlayer distance $d$ and embedded in a bulk dielectric medium representing the hBN encapsulation. In the case of a single layer this model leads to the standard Keldysh potential. We then solve the electrostatic problem to obtain an effective interaction between charge carriers within a given TMD layer. This simultaneously includes screening from the layer itself, from the opposite TMD layer, and from the encapsulation (see SI). Once the interaction is known, the intralayer exciton spectra are obtained by solving the corresponding Wannier equations. This requires a number of parameters that are separately constrained by experiments: the electron and hole masses, which we have taken from ARPES[32] and magneto-transport experiments[33,34]; the in- and out-of-plane dielectric constants of hBN, $\epsilon_\parallel = 6.9$ and $\epsilon_\perp = 3.7$[35,36]; and the interlayer separation $d = 6.47$ Å, taken as the average interlayer distance in bulk MoSe$_2$ and WSe$_2$[37,38]. The latter value lies at the bottom end of the range of layer step distances measured for MoSe$_2$ monolayers on WSe$_2$ by AFM[39,40]. As a result, the model contains only two free parameters, $\kappa_{MoSe_2}$ and $\kappa_{WSe_2}$, which we have constrained by six pieces of experimental data: the A2s-A1s energies for the two monolayers and the heterostructures reported here, as well as the A4s-A1s and A3s-A1s energies for monolayers reported by Liu et al.[21] and Chen et al.[23] (Table 3). Using $\kappa_{WSe_2} = 30.83$ Å and $\kappa_{MoSe_2} = 29.43$ Å, our theory reproduces all six exciton energy splittings to within 2.0%, all values being within twice the experimental error for these energies (Table 2). As shown in Table 2 these also fall within twice the estimated errors of the experimental values. Thus, whilst it is not obvious that the simplification of treating the layers as thin relative to their separation will work there is currently no need for a more complicated model.

It is worthwhile mentioning that further simplifications to the interaction model described above have a significant impact on its accuracy. For instance, treating the encapsulating medium as an isotropic dielectric with average dielectric constant of $\epsilon = \sqrt{\epsilon_\parallel \epsilon_\perp}$ increases the minimum error to 5%, or four times the experimental error for the six exciton energies. As discussed in the SI, in the limit of large separation between the charges the interaction takes the form of the Keldysh potential with the effective electric polarizability $\kappa = \kappa_{MoSe_2} + \kappa_{WSe_2} + \epsilon_\parallel d/2\pi$. This coincides with the long-range limit of interlayer interactions studied in[31], where it was found that this approximation underestimates screening at short distances. Here, we find that this approximation for the intralayer interaction has the opposite shortcoming, severely underestimating the strength of short range interactions by overestimating the screening contribution from the opposite layer, leading to errors of up to 30% when comparing to the measured spectra.

Based upon the agreement between theory and experiment we can now predict the excitonic binding contribution to the shift in the energies of the A1s excitons when forming the bilayers. The theory gives the difference in the binding energies of the A1s excitons in monolayers and the heterostructure are -8 ± 0.5 meV for $WSe_2$ and -19 ± 1 meV for $MoSe_2$. The small predicted errors are due to the fact that changes in the model's input parameters cause correlated changes in the monolayer and heterostructure energy predictions. In addition to changes in the excitonic binding energies there are two other contributions to the changes in the exciton energies on formation of the heterostructure; hybridisation of the excitons and changes in the single particle band gaps. Raman scattering has been shown to be sensitive to hybridisation of excitons[25] and there is no evidence for strong hybridisation of any of the excitons discussed in this paper. Therefore, we can conclude that the only other significant contribution to the absolute energies of the excitons is changes in the single particle state energies and therefore we find that the decrease in the single particle band gaps on formation of the heterostructure are 35 ± 2 meV for $MoSe_2$ and 46 ± 5 meV for $WSe_2$.

Whilst so far, we have focussed on the A excitons we also have results for the $MoSe_2$ B 1s and 2s excitons. In the $MoSe_2$ monolayer, we find a slightly larger 1s-2s separation for the B exciton (Table 2). This is consistent with ab initio predictions[41] of larger effective masses for both the B electron and hole. In the heterobilayer the $MoSe_2$ 1s-2s separation for the B exciton is 10-20 meV less than for the A excitons. It is not possible to explain this using the current theory without a reduction of the electron or hole masses associated with the B exciton of around 50%. There is no experimental or theoretical reason for believing such a reduction occurs in a heterostructure. Instead, it is likely that some other effect beyond the current theory is occurring. Due to the higher energy of the B excitons there may be more states available for hybridisation which could affect the energy of the 1s and 2s excitons differently. However, hybridisation is likely to be twist angle dependent and the B 1s-2s energy differences measured in the two samples are remarkably similar considering one is near 0° twist and one near 60° twist. Despite our inability to fully explain the B excitons in terms of geometry modifications of the Coulomb interaction it is remarkable that it is possible to explain the energy of the A excitons in the heterostructure and the monolayers using the same set of parameters for the dielectric and band structure properties and the separately constrained interlayer separation distance.

The results presented here are an important advance in underpinning our understanding of the Coulomb potential in the TMDs and other van der Waals structures. They present a method based upon resonance Raman scattering measurements of Raman peaks which are specific to 2s transitions which allows the energy and width of these transitions to be determined with useful precision. It is likely that the same method could be applied to other heterostructures containing at least one selenide layer and it may be possible to extend the method to the sulphides if 2s specific Raman peaks can be identified for these materials. Based upon these measurements we have shown that if we treat the screening due to the layers correctly it is possible to use a single set of parameters to predict the binding energies of both monolayer and heterobilayer excitonic states within 2.5%, i.e. to within current experimental errors. The availability of a reliable way to predict the binding energy of excitons in TMD heterostructures will accelerate the interpretation of a wide range of other experiments. For instance, we have used these predictions to determine the change in the single particle bandgaps for the two constituent layers. They will also be important to the quantification of the effect of hybridisation on excitons[42] and single particle states. This is particularly true for hybridisation effects that do not depend on twist angle, e.g. Förster energy transfer. As the TMD structures being produced increase in complexity we need to be able to quantitatively predict their properties with confidence and accuracy and this work is an important step in this process.

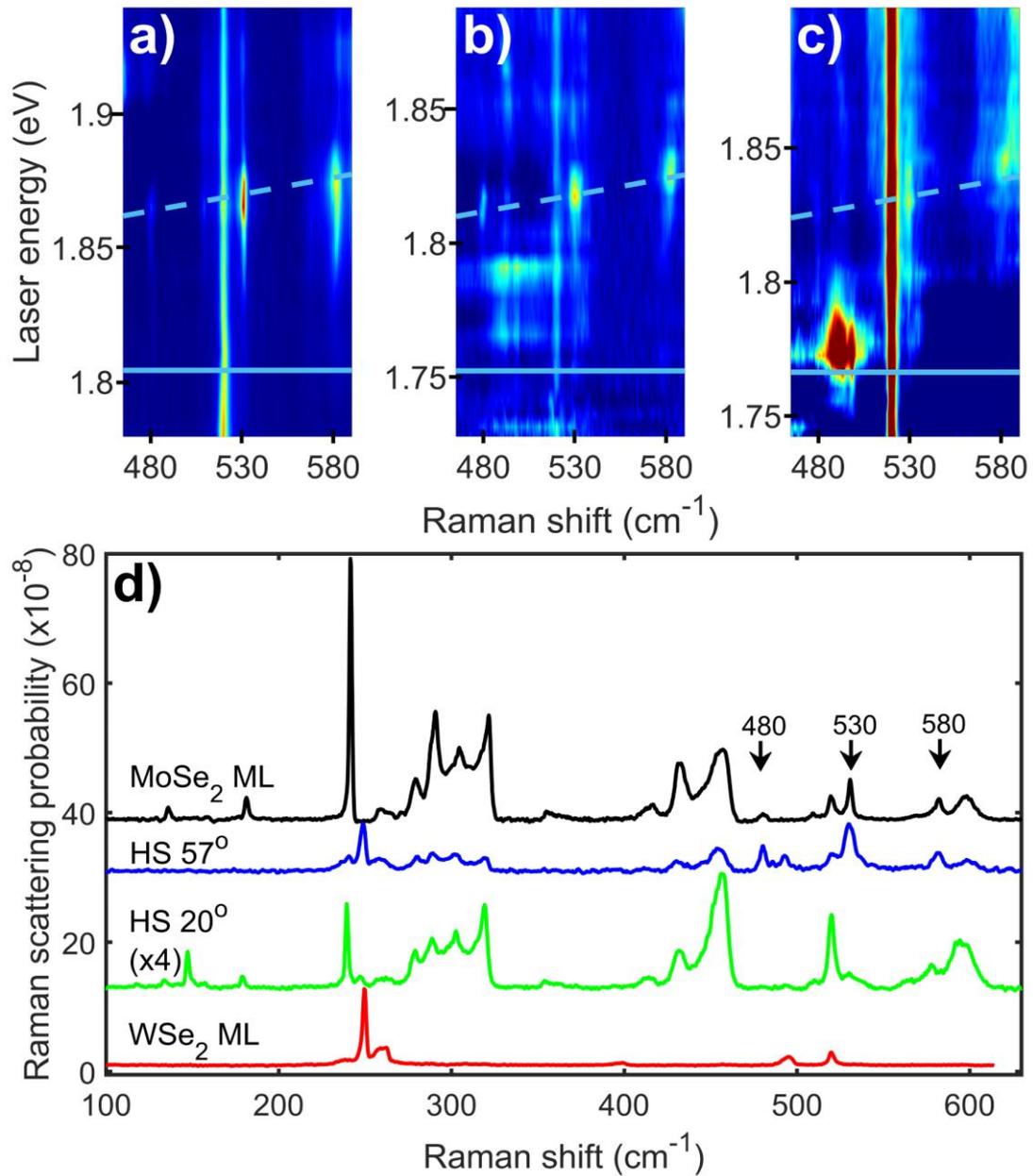

Figure 1: (a)-(c) Resonance Raman colour maps of a MoSe$_2$ (ML) (a) and two MoSe$_2$/WSe$_2$ heterostructures with twist angles of 57 ° (b) and 20 ° (c). The three key Raman peaks, that are only observed at the 2s resonances of MoSe$_2$ have their frequencies marked at 480, 530 and 580 cm$^{-1}$. The colour maps are presented with a logarithmic scale for the intensity to enhance the weaker peaks. The light blue lines represent the energies of the incoming (solid) and outgoing (dashed) resonances with the MoSe$_2$ intralayer A2s exciton (See S3 in the supplementary information for additional discussion of incoming/outgoing resonances). The energy of the exciton was determined by fitting the resonance behaviour of key Raman peaks and is presented in Table 1. (d) Raman spectra for the MoSe$_2$ (ML) and two MoSe$_2$/WSe$_2$ heterostructures plus a WSe$_2$ monolayer at the outgoing resonance energy of the A2s exciton, showing the key Raman peaks with their frequencies marked by arrows.

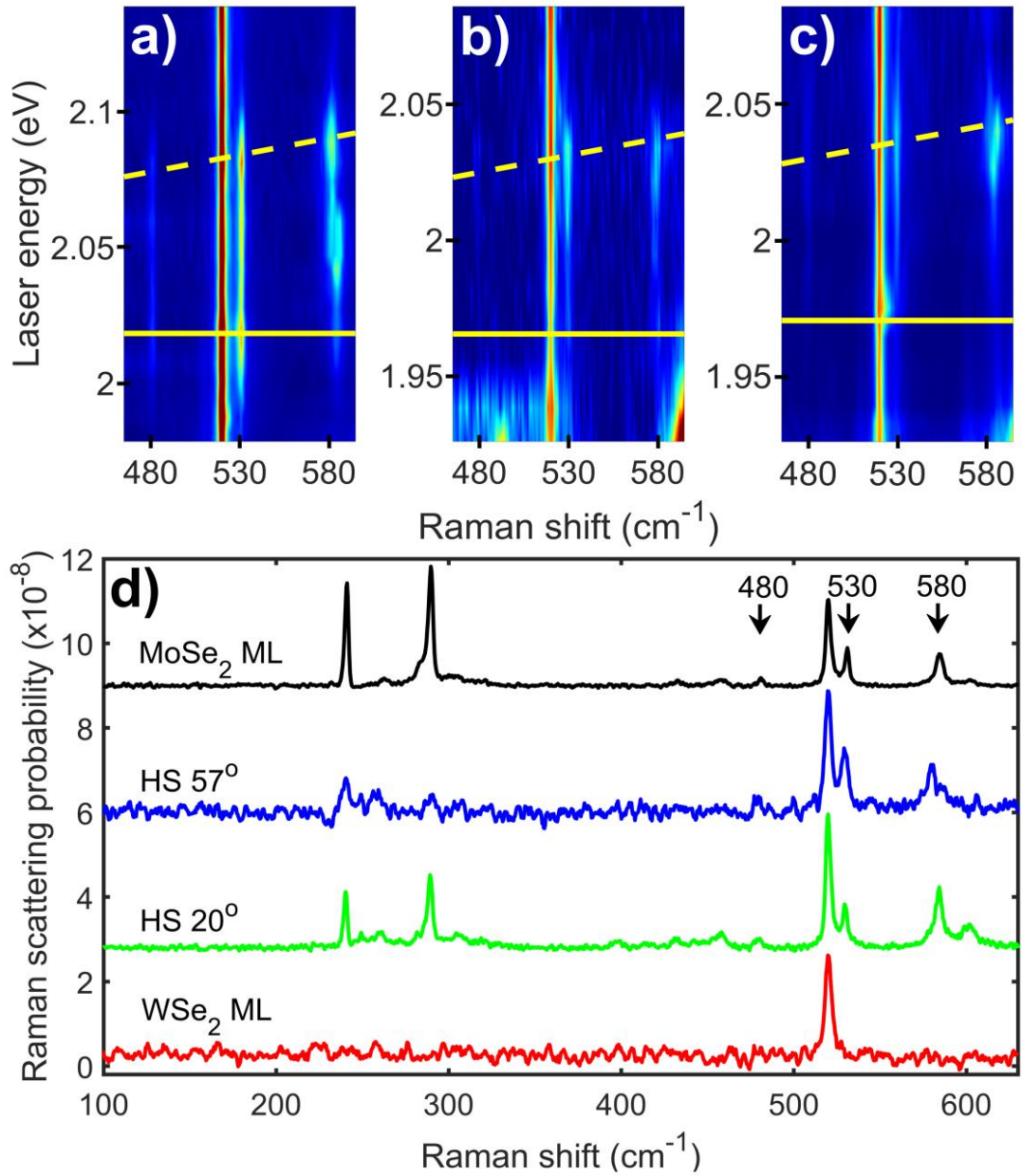

*Figure 2: (a)-(c) Resonance Raman colour maps of a $MoSe_2$ monolayer (ML) (a) and two $MoSe_2/WSe_2$ heterostructures with twist angles of 57 ° (b) and 20 ° (c) across the $MoSe_2$ B2s intralayer exciton resonance. The three key Raman peaks, that are only observed at the 2s resonances of $MoSe_2$ have their frequencies marked at 480, 530 and 580 $cm^{-1}$. The plots are presented with a logarithmic scale for intensity to enhance the weaker peaks. The yellow lines represent the energies of the incoming (solid) and outgoing (dashed) resonances with the $MoSe_2$ intralayer B2s exciton. The energy of the exciton was determined by fitting the resonance behaviour of the key Raman peaks and is presented in Table 1. (d) Raman spectra for the $MoSe_2$ monolayer (ML) and two $MoSe_2/WSe_2$ heterostructures plus a $WSe_2$ monolayer at the outgoing resonance of the $MoSe_2$ B2s exciton, showing the key Raman peaks with their frequencies marked by arrows.*

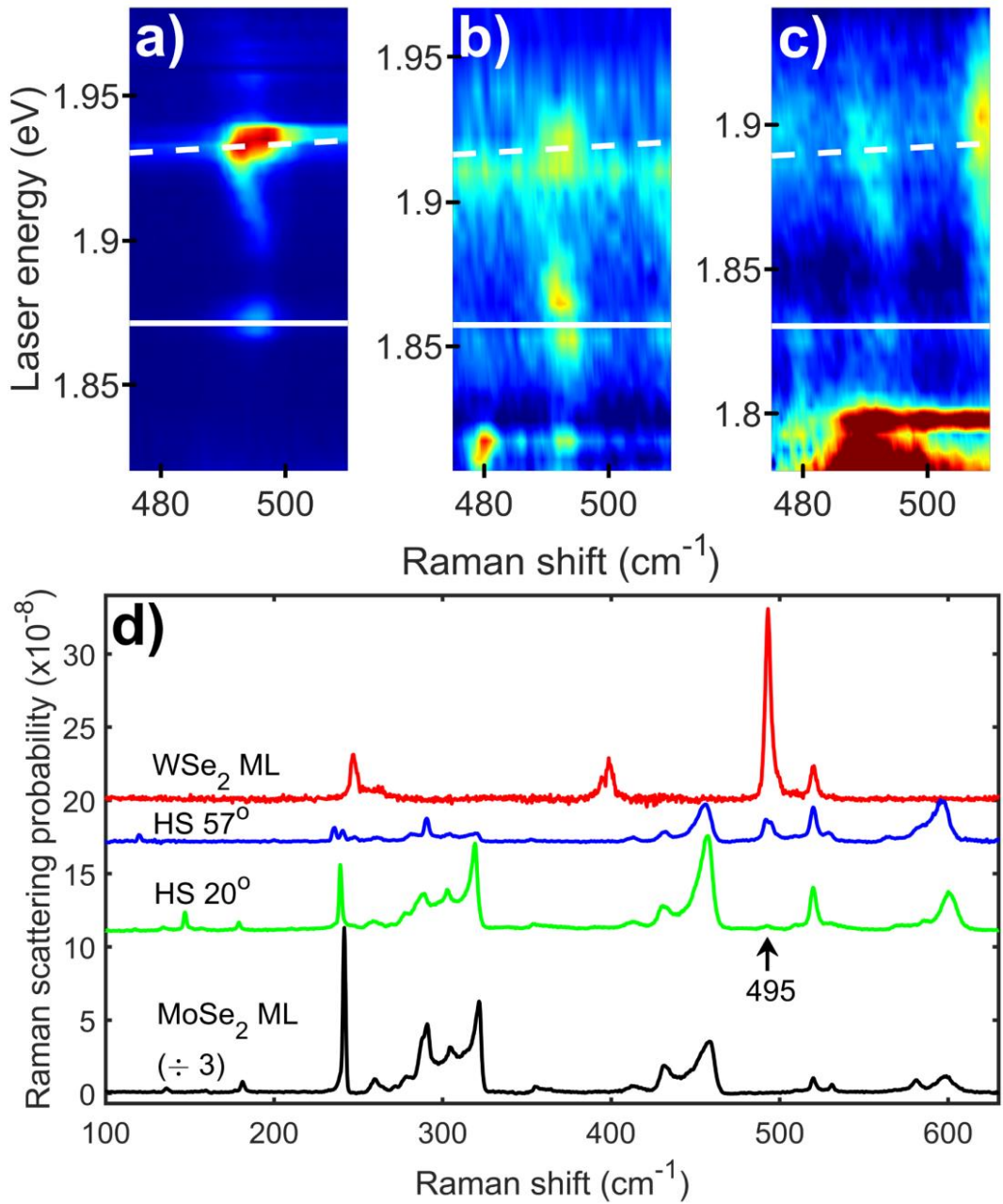

Figure 3: (a)-(c) Resonance Raman colour maps of a WSe$_2$ monolayer (ML) (a) and two MoSe$_2$/WSe$_2$ heterostructures with twist angles of 57 ° (b) and 20 ° (c) for excitation energies near the WSe$_2$ A2s intralayer exciton resonance. The key A2s specific Raman peak is at around 495 cm$^{-1}$. The plots are presented with a logarithmic scale for intensity to enhance the weaker peaks. The white lines represent the energies of the incoming (solid) and outgoing (dashed) resonances with the WSe$_2$ intralayer A2s exciton. The energy of the exciton was determined by fitting the resonance behaviour of the key Raman peak and is presented in Table 1. (d) Raman spectra for the WSe$_2$ monolayer (ML) and two MoSe$_2$/WSe$_2$ heterostructures plus a MoSe$_2$ monolayer at the outgoing resonance of the WSe$_2$ A2s exciton. The frequency of the associated 495 cm$^{-1}$ Raman peak is marked with an arrow.

|  |  | Measured Exciton Energy in Sample (eV) | | | |
| --- | --- | --- | --- | --- | --- |
| Material | Exciton | MoSe$_2$ ML | WSe$_2$ ML | HS (57°) | HS (20°) |
| MoSe$_2$ | A1s | 1.648 ± 0.001 | --- | 1.620 ± 0.001 | 1.623 ± 0.001 |
| MoSe$_2$ | A2s | 1.803 ± 0.002 | --- | 1.754 ± 0.001 | 1.765 ± 0.002 |
| MoSe$_2$ | B1s | 1.858 ± 0.001 | --- | 1.839 ± 0.001 | 1.849 ± 0.001 |
| MoSe$_2$ | B2s | 2.016 ± 0.001 | --- | 1.961 ± 0.016 | 1.969 ± 0.004 |
| WSe$_2$ | A1s | --- | 1.740 ± 0.001 | 1.728 ± 0.002 | 1.718 ± 0.002 |
| WSe$_2$ | A2s | --- | 1.871 ± 0.001 | 1.857 ± 0.003 | 1.835 ± 0.004 |
| WSe$_2$ | B1s | --- | 2.166 ± 0.001 | 2.161 ± 0.007 | 2.164 ± 0.001 |

Table 1: Energies of the various excitonic transitions obtained by fitting from the Raman results for hBN encapsulated vdW MoSe$_2$/WSe$_2$ heterostructure samples (HS) with twist angles of 57 and 20 ° and their constituent monolayers (ML).

|  |  | Energy differences for Exciton States (meV) | | | | | |
| --- | --- | --- | --- | --- | --- | --- | --- |
|  |  | MoSe$_2$ ML | | WSe$_2$ ML | | Heterostructures | | |
| Material | Exciton | Experiment | Theory | Experiment | Theory | (57°) | (20°) | Theory |
| MoSe$_2$ | A1s-A2s | 155 ± 3 | 152.3 | --- | --- | 134 ± 2 | 142 ± 3 | 138.2 |
| MoSe$_2$ | B1s-B2s | 158 ± 2 | --- | --- | --- | 122 ± 17 | 120 ± 3 | --- |
| MoSe$_2$ | A1s-B1s | 210 ± 2 | --- | --- | --- | 219 ± 2 | 226 ± 2 | --- |
| WSe$_2$ | A1s-A2s | --- | --- | 131 ± 1 | 130.2 | 129 ± 4 | 117 ± 4 | 122.4 |
| WSe$_2$ | A1s-B1s | --- | --- | 426 ± 2 | --- | 433 ± 8 | 446 ± 3 | --- |

Table 2: Energy separations between key excitonic transitions within each sample determined from the results in Table 1 plus predictions for these energies from the model described in the main body of the paper. The agreement between experiment and theory is to within twice the error estimates in all cases.

|  |  | Energy differences for Exciton States (meV) | |
| --- | --- | --- | --- |
| Material | Exciton | Literature Experimental Values | Theory |
| MoSe$_2$ ML | A1s-A2s | 151[19];   150[43] | 152.3 |
| WSe$_2$ ML | A1s-A2s | 131[23];   131[21];   130[18] | 130.2 |
| WSe$_2$ ML | A1s-A3s | 157[23];   152[21];   152[18] | 151.0 |
| WSe$_2$ ML | A1s-A4s | 161[23];   161[21] | 157.9 |

Table 3: Comparison of the difference in energy between different excitonic states in MoSe$_2$ and WSe$_2$ monolayers predicted using the theory discussed in the main body of the paper with previously published experimental values. There is excellent agreement which is within the experimental errors where these are available.

**Methods**

Monolayers of $WSe_2$ and $MoSe_2$ were mechanically exfoliated from bulk crystals and then stacked with layers of hexagonal boron nitride using dry transfer methods[44] onto a 290 nm $SiO_2$ coated Si wafer. All spectra in this paper were taken at 4 K (unless specified otherwise) with the samples in an Oxford Instruments High Resolution liquid helium flow Microstat. This was mounted on a 3-axis translation stage which allowed accurate positioning of the < 3 µm laser spot from the 50x Olympus 0.5 NA microscope objective on the samples. For the Raman spectra a CW Ti:Sapphire laser, provided excitation energies from 1.24 to 1.77 eV, and a dye laser using DCM, Rhodamine 6G and Rhodamine 110 laser dyes provided excitation energies from 1.74 to 2.25 eV. Reflectivity spectra were taking using a Fianium Supercontinuum laser. An Ocean optics HR4000 was used for the reflectivity spectra and a Princeton Instruments Tri-Vista triple spectrometer equipped with a liquid nitrogen cooled CCD was used for the Raman spectra. For each chosen wavelength, Raman spectra were taken with the spectrometer having both parallel linear polarization and crossed linear polarization to the laser. The Raman peaks were observed to have strongly co-linear polarization. The spectra were then subtracted, eliminating any non-Raman features such as photoluminescence - which can be observed when exciting just above the excitons. The constant 520 $cm^{-1}$ peak from the Si substrate was used as an internal reference which allowed us to calibrate the Raman shift of the TMD peaks to within 0.5 $cm^{-1}$. The intensities of the Raman peaks were also converted to absolute scattering probability using the Si peak intensity as a reference[28] with corrections for Fabry-Perot effects modelled using the reflectivity spectra as were performed in our previous paper[8]. The laser power incident on the sample was kept below 100 µW to avoid photo-doping and laser heating effects[45,46].

**Supporting Information**

The supplementary information contains reflectivity measurements on all 4 sample areas presented in the main body of the paper as well as more detailed discussion of the analysis of the resonance Raman data. Additional analysis of the $A_1'(\Gamma)$ and 2s associated Raman peaks in the monolayers and a discussion of the linewidths of the excitonic states is also included. Further details on the exciton energy modelling calculations are also included.

**Data Availability**

The data presented in this paper is openly available from the University of Southampton Repository with the DOI: https://doi.org/10.5258/SOTON/D1747

**Author Contributions**

The experiments were conceived by D.C.S, L.P.M and X.X. Samples were fabricated by P.R. The experimental measurements were performed by J.V and L.P.M. Data analysis and interpretation was carried out by J.V, D.C.S and L.P.M. The theoretical calculations were performed by D.A.R. The paper was written by D.C.S and J.V. All authors discussed the results and commented on the manuscript.

**Corresponding Author**

*D.C.Smith@soton.ac.uk

   Present Addresses

**Competing financial interests**

The authors declare no competing financial interests.

**Funding Sources**

This research was supported by UK Engineering and Physical Sciences Research Council via program grant EP/N035437/1 and grants EP/S019367/1, EP/S030719/1, EP/N010345/1, EP/V007033/1. VF acknowledges ERC Synergy Grant Hetero2D and EU Quantum Technology Flagship project 2D-SIPC. Both L.P.M and J.V were supported by EPSRC DTP funding. The work at U. Washington was funded by the Department of Energy, Basic Energy Sciences, Materials Sciences and Engineering Division (DE-SC0018171). D.R-T. was funded by UNAM-DGAPA. Work at Universidad Nacional Autónoma de México was funded by the DGAPA postdoctoral scholarship program and by the Sistema Nacional de Investigadores stimulus, CONACyT, México.

# Supplementary Information for Excited Rydberg States in TMD Heterostructures

*Jacob J.S. Viner [1], Liam P. McDonnell [1], David A. Ruiz-Tijerina[2], Pasqual Rivera [3], Xiaodong. Xu[3], Vladimir I. Fal'ko [4,5], David C. Smith [1*]*

1 School of Physics and Astronomy, University of Southampton, Southampton SO17 1BJ, United Kingdom.

2 Instituto de Física, Universidad Nacional Autónoma de México. Apartado Postal 20-364, Ciudad de México, 01000, México

3 Department of Physics, University of Washington, Seattle, WA, USA

[4]National Graphene Institute, University of Manchester, M13 9PL, United Kingdom.

[5] Henry Royce Institute for Advanced Materials, University of Manchester, Manchester, M13 9PL, United Kingdom.

**Table of Contents**

# S1) Reflectivity Spectra of TMD Monolayers and Heterostructures

Reflectivity spectra were measured for all sample areas relative to the $SiO_2$ coated Si substrate using a Fianium white light supercontinuum laser and Ocean Optics HR4000 fibre coupled spectrometer. These spectra were fitted to a T-Matrix reflectivity model which used refractive indices for the non-TMD layers reported in refs[S1–4]. The permittivity, $\varepsilon$, of the TMD layers was modelled as a sum of Lorentzian oscillators as in equation (1).

$$\varepsilon(E) = 1 + \sum_{k=1}^{n} \frac{a_k}{E_k^2 - E^2 - 2iE\Gamma_k} \quad (1)$$

where $a_k$, $E_k$ and $\Gamma_k$ are the amplitude, energy and width of the Lorentzian oscillator for excitonic state $k$, $E$ is photon energy and $n$ is the number of fitted states. Figure S1 shows the differential reflectivity for the $MoSe_2$ and $WSe_2$ monolayers and HS 57 ° and HS 20 ° and the resulting fits. The fitted features in the spectra corresponding to the 1s states are labelled and the spectra have arrows indicating the energies of the 2s states. In the monolayer cases, small bumps are visible in the spectra at these energies but they are not clear enough to stand out from the noise. Blue ($WSe_2$) and yellow ($MoSe_2$) arrows also mark the measured energies of the states in the two heterostructure spectra. No clear features were observed in the reflectivity spectra at these energies. The spectra also illustrate the overlapping energy of the easily visible $MoSe_2$ B1s state with the energy of the $WSe_2$ A2s indicated by a blue arrow.

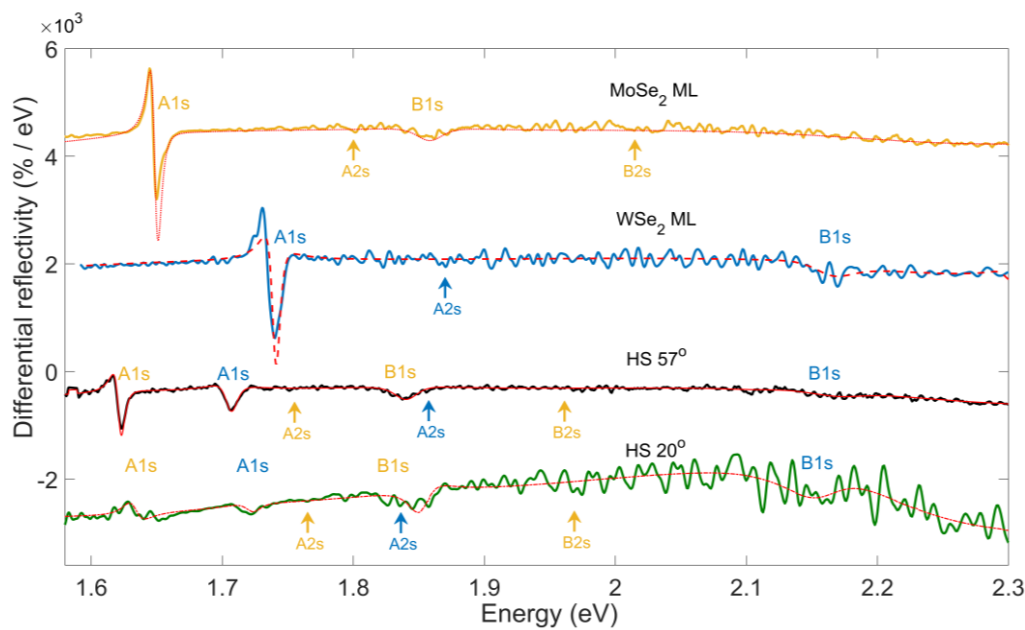

*Figure S1:* Differential reflectivity spectra are shown for the 57 and 20 ° heterobilayers along with spectra for both monolayer $MoSe_2$ and $WSe_2$. The red lines are the result of fitting the spectra using a summation of Lorentzian lineshapes to determine the linewidths and energies of the observed excitonic states. The $MoSe_2$ A1s and B1s excitons are labelled in yellow and the $WSe_2$ A1s and B2s excitons are labelled in blue. Yellow and blue arrows indicate the energies of the 2s states found by fitting the resonance Raman profiles for $MoSe_2$ and $WSe_2$ respectively.

## S2) Correction of Raman Spectra to Absolute Scattering Probability

For each laser energy, the Raman spectra are taken with the incoming beam being linearly polarised at a fixed angle. In each case a spectrum of the parallel and perpendicularly polarised scattered light is taken and these are subtracted to give only the co-polarised Raman signal. This removes the majority of luminescence from the spectra, which was the same intensity in the parallel and perpendicular spectra. This allowed for more reliable and accurate fitting of the Raman peaks in the spectra. In some cases, additional background subtraction is required if the luminescence is strong. This is due to small fluctuations in power of the laser over time between taking the parallel and perpendicular polarised spectra. The spectra were checked pre-subtraction for any Raman modes that were present in the perpendicularly polarised spectra but not the parallel spectra but no such peaks were found.

The Raman spectra all contain a 520 cm$^{-1}$ peak from the Si substrate with known resonance behaviour[S5]. For all laser energies, this peak is fitted and the spectra are normalised to the intensity of this fitted peak, and scaled using the absolute scattering probability of the Si peak[S5]. Finally, in order to correct for the optical cavity effects from the stacked layer structure, the spectra are corrected by the ratio of the Raman enhancement factors determined for the Si layer and TMD layer using:

$$R_{Layer} = \int_0^d |E_{Ab}(x) E_{Sc}(x)|^2 dx \qquad (2)$$

Where $R_{Layer}$ is the Raman enhancement factor for a layer in the stack of thickness d. $E_{Ab}$ and $E_{Sc}$ are the electric field strengths at the absorbed and scattered photon energies in that layer and x is the depth in the material from the interface. These electric field values come from the same T-Matrix code that models the reflectivity spectra in S2. This follows the approach by Yoon et. al.[S6].

## S3) Details of Fitting Procedure for Resonance Profiles

To extract the resonance profiles associated with each of the different Raman peaks the spectra were fitted to a summation of Lorentzian line shapes. Fitting the resonance profiles allows us to determine the energies and linewidths of the different excitonic states. For the simplest case where only one excitonic state is involved, a single state third order perturbation model can be used, given by equation (3):

$$I = \left| \frac{A}{(E - E_i - i\Gamma_i)(E - E_i - E_{p1} - i\Gamma_i)} \right|^2 \qquad (3)$$

Where $I$ is the Raman scattering intensity, $A$ is a constant representing the Raman scattering probability, $E$ is the laser (exciting photon) energy, $E_{p1}$ is the phonon energy $E_i$ is the energy of the excitonic state and $\Gamma_i$ is the linewidth of the excitonic state. For each phonon, there is both an incoming and outgoing resonance energy associated with an excitonic state. The incoming resonance corresponds to the case where the exciting photon energy matches the exciton energy. The outgoing resonance is then the case where the outgoing photon energy, equal to the incoming photon energy minus the phonon energy, matches the exciton energy. These cases correspond to the two parts of the denominator in equation (3). These incoming and outgoing resonances can be can be clearly separated, or overlap to create a single broad peak if the phonon energy is smaller than exciton linewidth.

However, this single state model cannot be used to describe more complex asymmetric resonances. For such resonances, a multiple excitonic state model can be used which includes interstate scattering terms[S7]. However, if the asymmetry is such that the one of the resonances cannot be observed, a simple Lorentzian fit is used, given by equation (4):

$$I = A^2 \frac{\Gamma_i^2}{(E - E_i')^2 + \Gamma_i^2} \quad (4)$$

Where $E_i'$ is the energy of the fitted resonance, $\Gamma_i$ is the width of the resonance, and $A$ is a constant representing the Raman scattering probability. This is used to extract the energy and width of an incoming ($E_i$) or outgoing resonance ($E_i + E_{p1}$). The A2s state peaks (480, 530, 580 for MoSe$_2$ and 495 for WSe$_2$) fall into the latter category in the heterostructures as they are weak or near strong resonances.

## S4) Comparison of the energies of the A2s and B2s exciton energies obtained by fitting A$_1'$(Γ) and the 2s-unique Raman peaks in monolayer MoSe$_2$

In the main body of the paper, we present the use of the 2s-unique Raman peaks at 480, 530 and 580 cm$^{-1}$ to determine the energy of the A2s and B2s excitons. In heterostructures it is not possible to use the 240 cm$^{-1}$ A$_1$'(Γ) peak as its behaviour is obscured by resonances with other states. Here we demonstrate that in ML the unique peaks yield the same resonance energy and width for the 2s state as the well understood 240 cm$^{-1}$ A$_1$'(Γ) in MoSe$_2$ monolayer.

In this monolayer the A2s resonance is close in energy to the much stronger B1s. As shown in fig S2, the effect of the A2s exciton on the resonance profile of the A$_1$'(Γ) is a smaller low energy shoulder on the B1s resonance. Fitting the combined resonance requires a two state scattering model as shown in Figure S2b). This model includes contributions from B1s-B1s and A2s-A2 scattering by the A$_1$'(Γ) phonon in addition from interstate phonon scattering; both B1s-A2s. This model is discussed in more detail in ref[S7].

Figure S3 shows the resonance profiles of the MoSe$_2$ monolayer A2s peaks with Raman shifts of 480, 530 and 580 cm$^{-1}$. For all of these three peaks, the same two state model is used to extract energies and widths for the A2s.

The energies obtained for the A2s based upon the fits to all four peaks are presented in Table S1. The results are all within experimental error of each other. The separation in energy we report between the A1s and A2s in monolayer MoSe$_2$ of 155 ± 3 meV falls within the range of previously published values for monolayers of between 150 – 168 meV[S8–11].

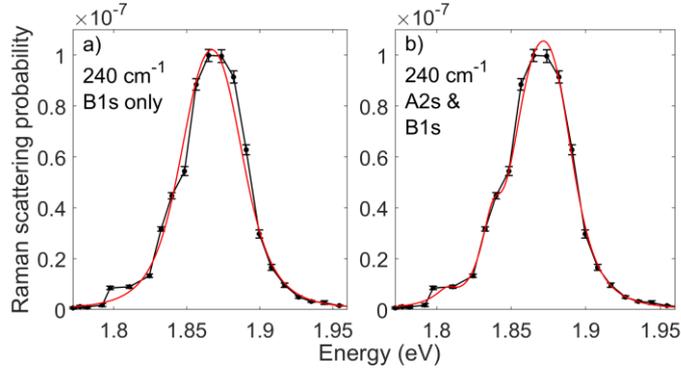

***Figure S2***: a) shows a single state fit to the 240 cm$^{-1}$ peak resonance profile in monolayer MoSe$_2$ at the B1s state, with the data showing a lower energy shoulder. b) Shows the 240 cm$^{-1}$ A$_1$'(Γ) peak fitted to a two excitonic state model with the two states being the MoSe$_2$ A2s and B1s states. The strongest contribution is from the higher energy B1s state and the lower energy shoulder is due to the A2s state. The resonance energy and width of the 2s state from this fit is given in Table S1.

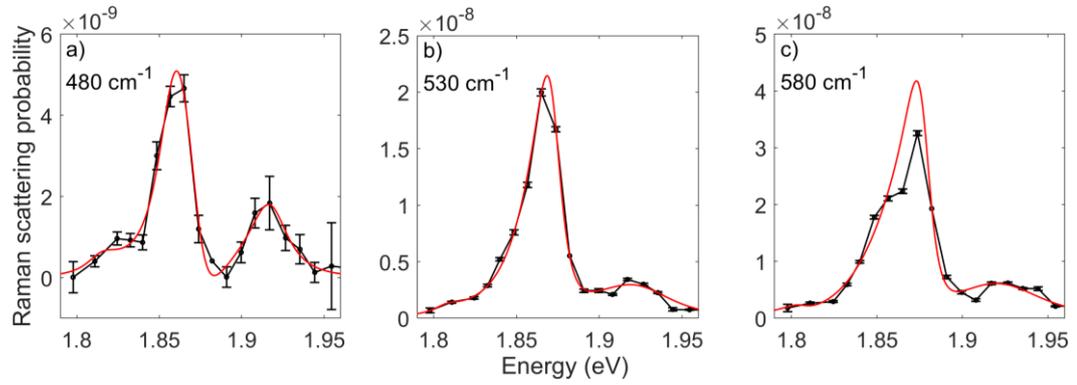

***Figure S3***: Fitted resonance profiles of monolayer MoSe$_2$. a)-c) show the three Raman peaks associated with the MoSe$_2$ A2s states at the outgoing resonance, fitted to 2 state scattering models. The A2s resonance energy and fitted widths are given in Table S1.

| Shift | Energy (eV) | Width (meV) |
|---|---|---|
| 240 | 1.803 ± 0.002 | 8 ± 1 |
| 480 | 1.813 ± 0.010 | 11 ± 3 |
| 530 | 1.806 ± 0.008 | 10 ± 1 |
| 580 | 1.809 ± 0.011 | 11 ± 2 |

***Table S1***: Parameters from fitting Raman peaks at the MoSe$_2$ A2s resonance in monolayer. The 480, 530 and 580 peaks and the 240 cm$^{-1}$ A$_1$'(Γ) peak was fitted to a two-state resonance model, where the other state present was the B1s. The errors given are standard errors from the fits which are also shown in Figure S3.

Unlike the A2s, the MoSe$_2$ B2s resonance (Fig S4) is isolated. In the case of the 240 cm$^{-1}$ peak, Figure S4 d), we observe this resonance on the tail of the much stronger B1s resonance. On the high energy side of the B2s resonance the scattering intensity of the 240cm$^{-1}$ peak does not fall to zero. This latter observation is probably due to the onset of resonance with the C exciton band or non-resonant scattering.

The resonance behaviour of the 240 cm$^{-1}$ peak, as well as the three 2s associated peaks with Raman shifts of 480, 530 and 580 cm$^{-1}$, are reasonably well fitted by a single state scattering model (equation 3). The incoming and outgoing resonances are distinguishable for the higher shift peaks but for the 240 cm$^{-1}$ peak they are close enough in energy to appear as one peak.

The best fit parameters are given in Table S2 and the energy of the resonance from the 240 cm$^{-1}$ peak falls within 1 meV of the average of the energies from the three 2s associated peaks and within 6 meV of three individual energies. Whilst the linewidths are not entirely consistent the two linewidths obtained from the best two fits, 530 and 480 cm$^{-1}$, are in close agreement. Comparing the value of the B1s-B2s exciton separation in monolayer of 158 ± 2 meV we find reasonable agreement with published values in literature around 160 meV[S8,10].

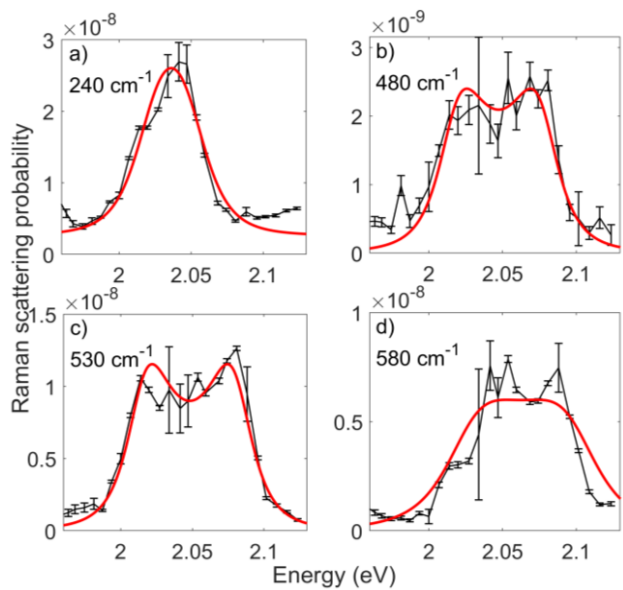

*Figure S4*: Resonance profiles for monolayer MoSe$_2$ Raman peaks at 580 (a), 530 (b), 480 (c) and 240 cm$^{-1}$ (d) when resonant with the MoSe$_2$ B2s exciton. The fitted amplitude is plotted in black with error bars showing the standard error of the fitted amplitude of the Raman peaks in the spectra. The red lines are from fits to a single state model (equation 3) and the fitting parameters used are presented in Table S2.

| Shift | Energy (eV) | Width (meV) |
|---|---|---|
| 240 | 2.021 ± 0.002 | 35 ± 4 |
| 480 | 2.018 ± 0.002 | 20 ± 3 |
| 530 | 2.015 ± 0.002 | 20 ± 2 |
| 580 | 2.027 ± 0.002 | 34 ± 11 |

*Table S2:* Energies and widths of the MoSe$_2$ B2s exciton from fitting resonance profiles of the $A_1'(\Gamma)$ and 2s associated Raman peaks in monolayer MoSe$_2$ to single state resonance models. The errors shown are the standard error given by the peak fitting process.

## S5) Comparison of the energies of the A2s exciton energies obtained by fitting A$_1'$(Γ) and the 2s-unique Raman peaks in monolayer WSe$_2$

In the main body of the paper we also used the 495 cm$^{-1}$ Raman peak, which in monolayer we only observe resonant with the A2s exciton, to identify the WSe$_2$ A2s exciton in the MoSe$_2$/WSe$_2$ heterostructure samples. It is not possible to use the main A1'(Γ) peak at 250 cm$^{-1}$ due to much brighter scattering from Raman peaks associated with the MoSe$_2$ B1s exciton, which covers a nearby energy range. Here we will show that the widths and energies of the exciton found from the 495 cm$^{-1}$ match the most studied 250 cm$^{-1}$ peak in monolayer WSe$_2$.

The fitted resonance profile of the WSe2 A$_1'$(Γ) peak at 250 cm$^{-1}$ in monolayer is presented in Figure S5 a). The incoming and outgoing resonances both clearly visible with similar amplitude. A single state resonance model (shown in red) is a good fit.

Figure S5 b) shows the 495 cm$^{-1}$ peak resonance profile over the same energy range. As well as being asymmetric, there is significant signal just below the outgoing resonance for which we do not have a clear explanation. In order to extract an energy and width for the resonance, we fit the strong outgoing resonance to a Lorentzian lineshape shown in red. As it is also visible in this case, the incoming resonance was fitted to a Lorentzian lineshape for comparison.

Table S3 shows the A2s energy and width of extracted from these fits to the 250 and 495 cm$^{-1}$ peaks. The three energies agree to within 1 meV and the widths by only slightly more, validating the use of the 495 cm$^{-1}$ peak as an identifier of the WSe$_2$ A2s state. This gives us an energy separation between the A1s and A2s excitons in monolayer of 131 ± 1 meV which agrees with the published values in literature of 130 - 133 meV[S9,12–16].

Figure S6 shows a colour map of the 495 cm$^{-1}$ peak at the A2s resonance in WSe$_2$ monolayer. The unusual resonance behaviour and dispersion suggests it involves intermediate electronic states at a saddle point in the exciton dispersion relation.

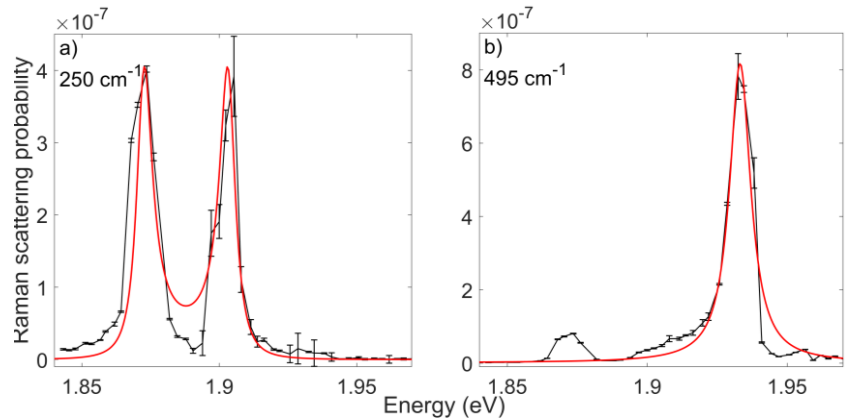

**Figure S5**: *Resonance profiles of the 250 (a) and 495 cm$^{-1}$ (b) Raman peaks in WSe$_2$ at the A2s resonance. The 250cm$^{-1}$ peak is fitted to a single state resonance model (equation 3) and the outgoing part of the 495 cm$^{-1}$ peak is fitted to a single Lorentzian lineshape. The extracted energies and widths from these fits are presented in Table S3.*

| Peak and Resonance | Energy (eV) | Width (meV) |
|---|---|---|
| 250 cm$^{-1}$ | 1.8714 ± 0.0005 | 3.5 ± 0.4 |
| 495 cm$^{-1}$ Incoming | 1.8711 ± 0.0005 | 3.6 ± 1.0 |
| 495 cm$^{-1}$ Outgoing | 1.8708 ± 0.0002 | 4.6 ± 0.3 |

**Table S3**: Comparison of energies found from fitting the incoming and outgoing resonance for the WSe$_2$ A2s 495 cm$^{-1}$ peak to a Lorentzian lineshape and fitting the 250 cm$^{-1}$ A1'(Γ) to a single state model. The errors given are standard error from the peak fit.

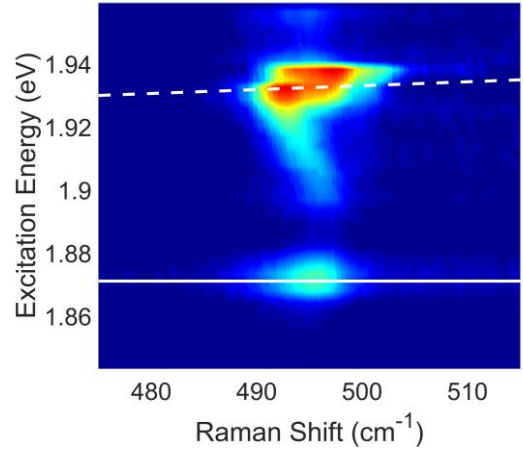

**Figure S6**: Colourmap of the WSe2 495 cm-1 peak at the A2s exciton resonance in monolayer. The white solid and dashed lines show the energies of the incoming and outgoing resonances respectively. The peak shows significant dispersion around 1.9 eV and above. It splits into 3 peaks at 1.93 eV which shift with energy independently of each other.

## S6) Comparison of Monolayer and Heterostructure Exciton Linewidths

Whilst not discussed in detail in the main body of the paper, the linewidths of the excitonic states were also extracted from the resonance Raman fits and are presented here in Table S4. Whilst the measured WSe$_2$ B1s linewidths were slightly smaller for the heterostructures, the larger uncertainties mean that the monolayer width still falls within the errors on the heterostructure values. Generally, the linewidths in the heterostructures are greater than those in the monolayers. This is due to the presence of lower energy interlayer excitons in the heterostructures with energies below the bright intralayer excitons[S17].

| | | Measured Exciton Width in Sample (meV) | | | |
|---|---|---|---|---|---|
| Material | Exciton | MoSe₂ ML | WSe₂ ML | HS (57°) | HS (20°) |
| MoSe₂ | A1s | 3.0 ± 0.1 | --- | 5.4 ± 0.3 | 15 ± 3.1 |
| MoSe₂ | A2s | 8.2 ± 0.7 | --- | 7.0 ± 1.7 | 13 ± 3.6 |
| MoSe₂ | B1s | 19 ±1.3 | --- | 23 ± 1.5 | 30 ± 1.6 |
| MoSe₂ | B2s | 9.4 ± 0.7 | --- | 13 ± 9.7 | 24 ± 7 |
| WSe₂ | A1s | --- | 5.6 ± 0.3 | 6.5 ± 1.1 | 10.8 ± 0.9 |
| WSe₂ | A2s | --- | 3.5 ± 0.2 | 10 ± 9.4 | 39 ± 14 |
| WSe₂ | B1s | --- | 32 ± 2.2 | 24 ± 11 | 29 ± 11 |

**Table S4**: Excitonic widths of the various excitonic transitions for monolayer (ML) and heterostructure samples (HS) with twist angles of 57 and 20 °.

## S7) Theoretical Model Details

Exciton states in a given TMD monolayer $\lambda = WSe_2, MoSe_2$ are solutions to the Wannier equation for the relative motion of the electron-hole system,

$$\left[\frac{\hbar^2}{2\mu^\lambda}\nabla_\rho^2 + U_\lambda(\rho)\right]\psi(\boldsymbol{\rho}) = E_b\psi(\boldsymbol{\rho}), \qquad (5)$$

where $\mu^\lambda = m_e^\lambda\, m_h^\lambda/(m_e^\lambda + m_h^\lambda)$ is the electron-hole reduced mass, $\boldsymbol{\rho} = \boldsymbol{r}_e - \boldsymbol{r}_h$ is the relative position vector, $U_\lambda(\rho)$ is the intralayer electron-hole interaction potential with $\rho \equiv |\boldsymbol{\rho}|$, and $E_b$ is the exciton binding energy.

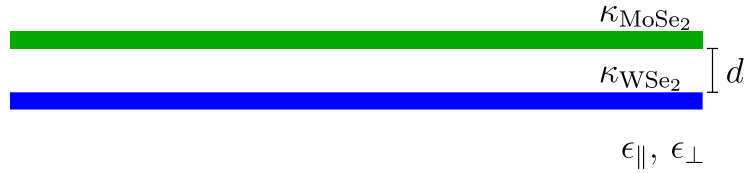

**Figure S7**: Dielectric model for the hBN-encapsulated WSe₂/MoSe₂ heterostructure. The MoSe₂ and WSe₂ layers are treated as infinitesimally thin dielectric films separated by a vertical distance d, with in-plane polarizabilities $\kappa_{MoSe_2}$ and $\kappa_{WSe_2}$, respectively. The hBN encapsulation enters our model as an environment with in- and out-of-plane dielectric constants $\epsilon_\parallel$ and $\epsilon_\perp$.

In the following, we treat TMD layers as infinitesimally thin dielectric films with dielectric polarizabilities $\kappa_{WSe_2}$ and $\kappa_{MoSe_2}$, embedded in a dielectric medium with in- and out-of-plane permittivities $\epsilon_\parallel$ and $\epsilon_\perp$ (Fig. S7).

### Excitons in monolayer WSe₂ and MoSe₂

For monolayers, the electron-hole interaction is described by the Rytova-Keldysh potential (in Gaussian units)[S18–20]

$$U_\lambda(\rho) = -\frac{\pi e^2}{2r_*^\lambda \tilde{\epsilon}}\left[H_0\left(\frac{\rho}{r_*^\lambda}\right) - Y_0\left(\frac{\rho}{r_*^\lambda}\right)\right]. \qquad (6)$$

Here, $\tilde{\epsilon}$ is the effective environment dielectric constant[S21]

$$\tilde{\epsilon} = \sqrt{\epsilon_\parallel \epsilon_\perp}, \tag{7}$$

$e$ is the (positive) fundamental charge, $r_*^\lambda = 2\pi\kappa_\lambda/\tilde{\epsilon}$ is the screening length of TMD $\lambda = WSe_2$ or $MoSe_2$, and $H_0$ and $Y_0$ are the zeroth Struve and Bessel functions of the second kind, respectively.

## Intralayer excitons in a WSe₂/MoSe₂ heterostructure

In a homo- or heterobilayer, the Coulomb interaction between charges inside a given layer is modified by additional screening coming from the opposite layer. Whereas this interaction does not have a closed-form expression as a function of $\rho$, its Fourier components are given by[S21]

$$U_{WSe_2}(q) = \frac{2\pi\left[1 + r_*^{MoSe_2} q\left(1 - e^{-2q\tilde{d}}\right)\right]}{\tilde{\epsilon}\left[1 + q\left(r_*^{WSe_2} + r_*^{MoSe_2}\right) + q^2 r_*^{WSe_2} r_*^{MoSe_2}\left(1 - e^{-2q\tilde{d}}\right)\right]}, \tag{8}$$

$$U_{MoSe_2}(q) = \frac{2\pi\left[1 + r_*^{WSe_2} q\left(1 - e^{-2q\tilde{d}}\right)\right]}{\tilde{\epsilon}\left[1 + q\left(r_*^{WSe_2} + r_*^{MoSe_2}\right) + q^2 r_*^{WSe_2} r_*^{MoSe_2}\left(1 - e^{-2q\tilde{d}}\right)\right]}, \tag{9}$$

with the renormalized interlayer distance

$$\tilde{d} = d\sqrt{\frac{\epsilon_\parallel}{\epsilon_\perp}}. \tag{10}$$

## Numerical solution to the Wannier equation

To solve Eq. (5) we use the method introduced by Griffin and Wheeler[S22], whereby the eigenvalue problem is evaluated in a finite *ad hoc* basis and numerically diagonalized. Since the potential is isotropic, we use the basis functions

$$\phi_{m,j}(\rho, \theta) = \frac{e^{im\theta}}{\sqrt{2\pi}} (\beta\rho)^{|m|} e^{-\beta_j \rho}, \tag{11}$$

where $\theta$ is polar angle of the vector $\boldsymbol{\rho}$, $m$ is the 2D angular momentum quantum number, and $\beta^{-1}$ is an arbitrary length scale used to make the basis functions dimensionless. Each function () has the correct asymptotic behavior expected from Eq. (5) at $\rho \to 0$ and $\rho \to \infty$, with the parameter $\beta_j$ acting as a variational inverse Bohr radius. The set of length scales $\beta_j^{-1}$ are chosen to span the range of possible Bohr radii that the excitons may take, and are discretized as $\beta_j = \beta_1 e^{\xi(j-1)}$, with $\xi = (N-1)^{-1}\log(\beta_N/\beta_1)$ and $\beta_N^{-1} \lesssim r_* \ll \beta_1^{-1}$. Then, we can write an approximate solution to Eq. (5) in the form

$$\psi_m(\rho, \theta) = \sum_{j=1}^N A_{m,j} \phi_{m,j}(\rho, \theta). \tag{12}$$

Substituting $\psi_m$ into Eq. (5), left-multiplying by $\psi_m^*$ and integrating gives the generalized eigenvalue problem

$$[H_m - E_b S_m]A_m = 0, \tag{13}$$

where we have defined the column vector of coefficients $A_m = (A_{m,1}, \cdots, A_{m,N})^T$, the overlap matrix $S_m$ with elements

$$S_{m,\mu\nu} = \beta^{2|m|} \frac{\Gamma(2|m|+2)}{(\beta_\mu + \beta_\nu)^{2|m|+2}}, \tag{14}$$

and the Hamiltonian matrix $H_m = K_m + U_m$, with kinetic energy elements

$$K_{m,\mu\nu} = \frac{\hbar^2}{2\mu}\left[-(2|m|+1)\frac{\beta^{2|m|}\beta_\mu \Gamma(2|m|+1)}{(\beta_\mu + \beta_\nu)^{2|m|+1}} + \frac{\beta^{2|m|}\beta_\mu^2 \Gamma(2|m|+2)}{(\beta_\mu + \beta_\nu)^{2|m|+2}}\right], \tag{15}$$

and potential energy terms

$$U_{m,\mu\nu} = -\int_0^\infty d\rho\, \rho\, \phi_{m,\mu}^*(\rho) U_\lambda(\rho) \phi_{m,\nu}(\rho). \tag{16}$$

The interaction form depends on whether the potential $U_\lambda$ corresponds to the monolayer case, Eq. (6), or to the heterobilayer case, Eqs. (8) and (9). For a monolayer, the integral can be evaluated analytically in real space and reads

$$U_{m,\mu\nu}^{ML} = -\frac{e^2 \beta^{2|m|}}{(r_*^\lambda)^2\, \tilde{\epsilon}} \left[ \frac{\Gamma(2|m|+3)}{(\beta_\mu + \beta_\nu)^{2|m|+3}} F_{3,2}\left(|m|+1, |m|+\tfrac{3}{2}, |m|+2; \tfrac{3}{2}, \tfrac{3}{2}; -\frac{(r_*^\lambda)^{-2}}{(\beta_\mu+\beta_\nu)^2}\right) \right.$$
$$\left. - 4^{|m|}(r_*^\lambda)^{2|m|+3} \cos(m\pi)\, \Gamma^2(|m|+1) F_{2,1}\left(|m|+1, |m|+1; \tfrac{1}{2}; -r_*^2(\beta_\mu + \beta_\nu)^2\right)\right].$$

Here, $\Gamma(n)$ is the Gamma function and $F_{p,q}(a_1, \ldots, a_p; b_1, \ldots, b_q; x)$ is the generalized hypergeometric function.

In the heterostructure case we may develop the integral as

$$U_{m,\mu\nu}^{HS} = -\beta^{2|m|}\int_0^\infty d\rho\, \rho f_{\mu\nu}(\rho) U_\lambda(\rho) = -\frac{\beta^{2|m|}}{2\pi}\int d^2\rho f_{\mu\nu}(\rho) U_\lambda(\rho), \tag{17}$$

with the definition $f_{\mu\nu}(\rho) \equiv \rho^{2|m|} e^{-(\beta_\mu+\beta_\nu)\rho}$. Then, we take the Fourier transforms

$$f_{\mu\nu}(\rho) = \int d^2q\, e^{iq\cdot\rho} f_{\mu\nu}(q),$$

$$U_\lambda(\rho) = \int d^2q\, e^{iq\cdot\rho} U_\lambda(q),$$

where $U_\lambda(q)$ are given by Eqs. (8) and (9), and

$$f_{\mu\nu}(q) = \frac{2\pi}{(\beta_\mu + \beta_\nu)^{2|m|+2}}\Gamma(2|m|+2)\,_2F_1\left(|m|+1, |m|+\tfrac{3}{2}; 1; -q^2(\beta_\mu+\beta_\nu)^{-2}\right). \tag{18}$$

Substituting into (17) we get

$$U_{m,\mu\nu}^{HS} = -\frac{\beta^{2|m|}}{(2\pi)^2}\int_0^\infty dq\, q\, f_{\mu\nu}(-q) U_\lambda(q). \tag{19}$$

The integrand of (19) is well behaved, and the integral can be evaluated numerically.

For each angular momentum quantum number $m$ we set a basis of $N$ functions, $\{\phi_{m,1}, ..., \phi_{m,N}\}$, and evaluate all matrix elements (14), (15) and (16) for the corresponding layer. For sufficiently large $N$, the low-energy spectrum of Eq. (13) saturates to a set of energies, giving a good approximation to the low-lying exciton states of Eq. (5). We have verified this for the monolayer case by comparing with finite elements calculations. In practice, $N \approx 30$ basis functions are sufficient for the first 5 energy eigenvalues to saturate to within a fraction of 1 meV.

## Model parameters

Computing the exciton spectrum of Eq. (5) requires the following parameters: the electron and hole masses $m_e^\lambda$ and $m_h^\lambda$; the environment dielectric parameters $\epsilon_\parallel$ and $\epsilon_\perp$; the interlayer distance $d$; and the layer's screening lengths $r_*^\lambda$. Table S5 compiles values for the former five parameters obtained from the experimental and *ab initio* literature.

|  | WSe$_2$ | MoSe$_2$ | WSe$_2$/MoSe$_2$ | hBN |
|---|---|---|---|---|
| $m_e$ ($m_0$) | 0.50[a] | 0.80[a] | - | - |
| $m_h$ ($m_0$) | 0.42[b] | 0.50[b] | - | - |
| $d$ (Å) | - | - | 6.47[c,d] | - |
| $\epsilon_\parallel$ | - | - | - | 6.9[e,f] |
| $\epsilon_\perp$ | - | - | - | 3.7[e,f] |

**Table S5:** *Material parameters appearing in the Wannier equation (5), compiled from the literature. The WSe$_2$ and MoSe$_2$ electron (hole) masses are experimental values based on ARPES (magnetotransport) measurements. The interlayer distance for the WSe$_2$/MoSe$_2$ heterostructure was estimated as the average between the interlayer distances of bulk WSe$_2$ and bulk MoSe$_2$. The dielectric constants for hBN are based on optical spectroscopy measurements and ab initio calculations. References: a) Gustafsson et al. (2018)[S23]; b) Nguyen et al. (2019)[S24]; c) Al-Hilli and Evans (1972)[S25]; d) Hicks (1964)[S26]; e) Geick, Perry and Rupprecht (1966)[S27]; f) Laturia, Van de Put and Vandenberghe (2018)[S28].*

The WSe$_2$ and MoSe$_2$ screening lengths are kept as two free parameters to fit the experimental data. Combining our monolayer measurements for the A1s-A2s energy spacing in both materials (Table S4 and main text Table 2) with the results of Liu et al.[S15] and Chen et al.[S16] for the A3s-A1s and A4s-A1s energy spacings in WSe$_2$ (main text Table 3) gives 3 data points. In addition, since our model (5) cannot distinguish between the 57° and 20° heterostructure configurations, we averaged the A2s-A1s energy separation of each material over both samples (Table S4 and main text Table 2) to obtain two additional data points. In summary, we are fitting 5 data points with only 2 free parameters.

We computed the intralayer A-exciton $s$-state ($m = 0$) spectra for WSe$_2$ and MoSe$_2$, both in the monolayer and heterostructure configurations, for a range of values of $r_*^{WSe_2}$ and $r_*^{MoSe_2}$. Then, we chose the pair ($r_*^{WSe_2}, r_*^{MoSe_2}$) that simultaneously minimized the percentile errors

$$A_{\text{err}} = \frac{|A_{\text{experiment}} - A_{\text{theory}}|}{|A_{\text{experiment}}|} \cdot 100\%, \qquad (20)$$

for all five data points. The obtained values are $r_*^{WSe_2} = 38.34$ Å and $r_*^{MoSe_2} = 36.60$ Å, giving the percentile errors reported in Table S6, all of which are $\leq 2\%$. These screening lengths correspond to in-plane polarizabilities $\kappa_{WSe_2} = 30.83$ Å and $\kappa_{MoSe_2} = 29.43$ Å.

|  | Percentile error |
|---|---|
| Monolayer WSe$_2$ A2s-A1s | 0.60 |
| Monolayer WSe$_2$ A3s-A1s | 1.93 |
| Monolayer WSe$_2$ A4s-A1s | 1.93 |
| Monolayer MoSe$_2$ A2s-A1s | 0.20 |
| WSe$_2$ A2s-A1s in WSe$_2$/MoSe$_2$ | 0.48 |
| MoSe$_2$ A2s-A1s in WSe$_2$/MoSe$_2$ | 0.11 |

***Table S6***: Percentile error associated of the theoretical exciton energy spectra obtained from numerically solving Eq. (50), compared to the experimental values of this work, and of Liu et al.[S15] and Chen et al.[S16] Parameters: $r_*^{WSe_2} = 38.34$ Å and $r_*^{MoSe_2} = 36.60$ Å.